\documentclass{elsart}
\usepackage{graphics}
\usepackage{epsfig}
\usepackage{amssymb}
\begin{document}
\begin{frontmatter}

\title{Green's function theory for a magnetic impurity layer in a ferromagnetic
Ising film with transverse field}

\author{R. V. Leite},
\ead{valmir@fisica.ufc.br}
\author{B. T. F. Morais},
\author{J. Milton Pereira Jr.},
\ead{pereira@fisica.ufc.br}
\author{and R. N. Costa Filho}
\ead{rai@fisica.ufc.br}

\address{Departamento de F\'{\i}sica, Universidade Federal do
Cear\'a, CXP $6030$ Campus do Pici, $60451-970$ Fortaleza,
Cear\'a, Brazil}


\begin{abstract}
A Green's function formalism is used to calculate the spectrum of
localized modes of an impurity layer implanted within a
ferromagnetic thin film. The equations of motion for the Green's
functions are determined in the framework of the Ising model in a
transverse field. We show that depending on the thickness,
exchange and effective field parameters, there is a ``crossover''
effect between the surface modes and impurity localized modes. For
thicker films the results show that the degeneracy of the surface
modes can be lifted by the presence of an impurity layer.
\end{abstract}

\end{frontmatter}

\section{Introduction}

The spectrum of elementary excitations in solids is known to be
modified by the presence of surfaces in the medium in which they
propagate. This occurs because an interface breaks the
translational symmetry of the medium and can also modify the
microscopic scheme of interactions due to surface reconstruction
effects. The modifications of the spectrum can be signaled by the
appearance of surface localized modes, besides the volume
excitations. The translational symmetry of a crystal can also be
broken by the presence of defects or impurities in the lattice,
and localized excitations may result.

Localized excitations of magnetic media have been widely studied,
theoretically as well as experimentally \cite{MikeTilley} and
several models have been proposed to elucidate the dynamics of
surface modes in a variety of magnetic systems. Among these
models, the transverse Ising model has been shown to give a good
theoretical description of real materials with anisotropic
exchange (e.g., CoCs$_{3}$Cl$_{5}$ and DyPO$_{4}$), of materials
in which the crystal field ground state is a singlet \cite{Wong},
and has also been employed as a pseudo-spin model for
hydrogen-bonded ferroelectric materials \cite{Blinc}. Previous
calculations have obtained the spin wave (SW) spectrum of Ising
ferromagnets in a transverse external field for semi-infinite
systems \cite{Shiwai} and films \cite{Kontos}. These results were
obtained for ideal media, i.e. structures without defects or
impurities. Studies \cite{Wax,Dewames,Cottam,Tilley,Costa}
concerning excitations of semi-infinite systems described by the
transverse Ising model have predicted the existence of localized
SW modes, as well as bulk SW.

Within the quasi-particle approximation, the effect of a localized perturbation
in a periodic structure can be worked out exactly using Green's
functions techniques. These techniques have been
extensively used to calculate the impurity modes in
infinite and semi-infinite Heisenberg ferromagnets and
antiferromagnets \cite{Cowley,Chen}.

The purpose of this paper is to present the first calculation of
the SW spectrum of a thin ferromagnetic film with an impurity
layer in the framework of an Ising model in a transverse field.
Dispersion relations for the bulk, surface and impurity SW modes
are obtained by finding the poles of the Green's functions
associated with the spin operators of each layer in the film. This
calculation generalizes a previous study of the spin dynamics of
an impurity layer in a semi-infinite ferromagnet, presented in
Ref. \cite{Costa}.

\section{Green's function formalism}
Let us consider a retarded Green's function, defined in terms of
two time-dependent operators $A(t)$ and $B(t)$, in the
Heisenberg picture, as:
\begin{equation}
G_{r}(t-t^{\prime })= \langle \langle A(t);B(t^{\prime })\rangle
\rangle,
\end{equation}
which, in frequency representation, can be shown to satisfy the
equation of motion \cite{Zubarev},
\begin{equation}
\omega \left\langle \left\langle A;B\right\rangle \right\rangle
_{\omega }=\frac{1}{2\pi }\left\langle \left[ A,B\right]
\right\rangle +\left\langle \left\langle \left[ A,{\mathcal H}\right]
;B\right\rangle \right\rangle _{\omega },
\end{equation}
where $\omega $ is a frequency label, and ${\mathcal H}$
is the Hamiltonian of the system. In the present case we employ the
Ising Hamiltonian in a transverse field
\begin{equation}
{\mathcal H}=-\frac{1}{2}
\sum_{<i,j>} J_{ij}S_{i}^{z}S_{j}^{z}-h\sum_i S_{i}^{x},
\label{hamilt}
\end{equation}
where $S_{i}^{\alpha }(\alpha =x,y,z)$ is a spin operator
component at site $i$, $h$ is the applied transverse magnetic
field, taken as $h_S$ in the surface and $h_I$ in the impurity
layer, and $\ J_{ij}$ is the exchange coupling between
nearest-neighbor sites $i$ and $j$. The exchange parameter is
taken as $J$ in the bulk region, as $J_S$ in the surface and as
$J_I$ in the impurity layer. The exchange coupling between a spin
in the surface layer or in the impurity layer and a spin in the
adjacent bulk layers is assumed to be $J'_S$ and $J'_I$,
respectively. In the special case where the impurity layer and the
surface are adjacent we use $J_{SI}$ as the exchange interaction.
Next, we construct the equation-of-motion
\begin{eqnarray}
\omega \left\langle \left\langle S_{l}^{\alpha };S_{m}^{\beta
}\right\rangle \right\rangle _{\omega }=&\frac{1}{2\pi }
\left\langle \left[ S_{l}^{\alpha },S_{m}^{\beta }\right]
\right\rangle + \qquad \qquad  \\ \nonumber &\left\langle
\left\langle \left[ S_{l}^{\alpha },{\mathcal H}\right]
;S_{m}^{\beta }\right\rangle \right\rangle _{\omega }. \label{eqm}
\end{eqnarray}
where $l$ and $m$ are labels assigned to each lattice
site. Thus we obtain an infinite chain of coupled equations, which can then
be decoupled using the Random Phase Approximation \cite{Keffer},
\begin{eqnarray}
\left\langle \left\langle S_{l}^{\alpha }S_{j}^{\gamma
};S_{m}^{\beta }\right\rangle \right\rangle _{\omega }\approx &
\left\langle S_{l}^{\alpha }\right\rangle \left\langle
\left\langle S_{j}^{\gamma };S_{m}^{\beta }\right\rangle
\right\rangle _{\omega } +  \qquad \qquad  \\ \nonumber
&\left\langle S_{j}^{\gamma }\right\rangle
\left\langle \left\langle S_{l}^{\alpha };S_{m}^{\beta
}\right\rangle \right\rangle _{\omega },
\label{rpa}
\end{eqnarray}
where $l\neq j$. Using this approximation and inserting the Hamiltonian
Eq. (\ref{hamilt}) into the equation of motion Eq. (\ref{eqm}), we obtain
\begin{eqnarray}
A(\omega)\langle\langle S_{l}^{z};S_{m}^{z}\rangle \rangle
_{\omega }=&
h R_l^x \frac{\delta _{lm}}{2\pi}-\qquad \qquad \qquad \qquad \\ \nonumber
&\qquad h R_l^x\sum_jJ_{lj}\left\langle \left\langle
S_{j}^{z};S_{m}^{z}\right\rangle \right\rangle _{\omega },
\end{eqnarray}
where
\begin{eqnarray}
A(\omega)=\left\{\omega ^{2}-\sum_jJ_{lj}R_j^z
\left(\sum_iJ_{il}R_i^z\right)-h^{2}\right\}, \nonumber
\end{eqnarray}
with $R_l^z=\left\langle S_l^z \right\rangle$ and
$R_l^x=\left\langle S_l^x \right\rangle$.

Equation (6) may be solved to obtain the spin-dependent Green's
functions at any temperature. In this work we focus on the
paramagnetic phase, at $T\geq T_{c}$, where $T_c$ is the Curie
temperature of the ferromagnet. In this regime, due to the absence
of net magnetization, we have $\left\langle
S_{j}^{z}\right\rangle=0$. For spin S=1/2, we use the static spin
average in the $x$ direction calculated from the mean-field
theory\cite{Blinc}
\begin{equation}
R^{x}_\ell =\frac{1}{2}\tanh
\left(\frac{h_\ell }{2k_{B}T}\right),
\end{equation}
where, depending on the site layer , the index $\ell$ stands for $S,B,I$, for surface,
bulk, and impurity layers, respectively and $k_{B}$ is Boltzmann's constant.
After some straightforward algebra, the
Green's function is found as:
\begin{equation}
G_{nn\prime }=\frac{1}{%
2\pi }\left[\frac{hR^{x}_\ell }{\omega ^{2}-h^{2}+4hR^{x}_\ell J\gamma
({\bf{q}_\parallel})}\right],
\end{equation}
where $n$ and $n'$ are layer indices and $\gamma ({\bf{q}_\parallel})$ is the
structure factor, given by,
\begin{equation}
\gamma ({\bf{q}_\parallel})=\frac{1}{2}\bigl[\cos (q_{x}a)+\cos (q_{y}a)
\bigr].
\end{equation}

Some mathematical techniques, including general recursive
algorithms for layered systems, are available to solve these
equations. Here we obtain explicit results using an approach
analogous to earlier calculations for pure ferromagnets
\cite{Cottam}. We write the coupled equations in a matrix form
\begin{equation}
({\bf A}_{0}+{\bf \Delta} ){\bf G} = -\Bigl( \frac{1}{2\pi J}\Bigr)\bf{I}
\end{equation}
where ${\bf A}_0$ is a $N \times N$ tridiagonal matrix, with
$N$ being the total number of layers in the film. The elements of ${\bf A}_0$
are defined by
\begin{equation}
[A_0]_{l,m}=d\;\delta_{l,m}-(\delta_{l-1,m}+\delta_{l+1,m}),
\end{equation}
which represents the system without impurities,
and ${\bf \Delta}$ is a  $N \times N$ matrix containing the information
regarding the perturbing effects of
both the surfaces and the impurities, with elements given by
\begin{eqnarray}
&\Delta_{1,1}=\Delta_{N,N}=\Delta_S \\ \nonumber
&\Delta_{1,2}=\Delta_{2,1}=\Delta_{N,N-1}=\Delta_{N-1,N}=\Delta_{SB} \\ \nonumber
&\Delta_{l,l}=\Delta_I \\ \nonumber
&\Delta_{l+1,l}=\Delta_{l-1,l}=\Delta_{l,l+1}=\Delta_{l,l-1}=\Delta_{IB},
\end{eqnarray}
where the index $l$ refers to the impurity layer, and the other elements are
set to zero. The parameter $d$ is defined by:
\begin{equation}
 d=-\frac{\omega ^{2}-h^{2}+4hR^{x}_BJ\gamma
({\bf{q}_\parallel})}{hR^{x}_BJ} .
\label{diag}
\end{equation}

The parameters $\Delta_S$, $\Delta_{SB}$, $\Delta_I$ and
$\Delta_{IB}$, which depend on the surface and on the impurity are
\begin{equation}
\Delta_{S}=-\frac{\omega^{2}-h_{S}^{2}+4h_{S}R_{S}^{x} J_{S}\gamma
({\bf{q}_\parallel})}{h_{S}R_{S}^{x}J}-d
\end{equation}

\begin{equation}
\Delta_{SB}=-\frac{J-J_{S}}{J}
\end{equation}

\begin{equation}
 \Delta_{I}=-\frac{\omega
^{2}-h_{I}^{2}+4h_{I}R_{I}^{x}J_{I}\gamma ({\bf{q}_\parallel})}{%
h_{I}R_{I}^{x}J}-d
\end{equation}
and,

\begin{equation}
 \Delta_{IB}=-\frac{J-J_{I}^{\prime }}{J}.
\end{equation}

Finally, we rewrite the Green's function in
matrix form:
\begin{equation}
{\bf G}=-\left(\frac{1}{2\pi J}\right)\left[({\bf I}+{\bf
A}_0^{-1}{\bf\Delta})^{-1}\right]{\bf A}_0^{-1}.
\label{gfm}
\end{equation}
The inverse of the tridiagonal matrix ${\bf A}_0$ can be found in the
literature.

\section{Surface and impurity modes}

In order to calculate the dispersion relations of the localized SW
modes, one has to find the poles of the Green's function
Eq.(\ref{gfm}). These are obtained by solving the determinantal
equation \cite{Cottam}
\begin{equation}
\det({\bf I}+{{\bf A}_0}^{-1}{\bf \Delta})=0.
\label{det}
\end{equation}
The elements of the inverse matrix ${{\bf A}_0}^{-1}$ involve a factor $x$, which
is defined in terms of the diagonal element of ${\bf A}_0$ as
\begin{equation}
d=x+x^{-1}.
\end{equation}

Solutions of Eq. (\ref{det})  involving $|x|=1$ give us the
frequencies of the quantized bulk modes. Such solutions correspond
to $x=\exp(i\alpha a)$, where $\alpha$ is a set of real discrete
values $(0\leq \alpha\leq \pi)$ which depends on the film
thickness and can be obtained from the determinantal equation.
Therefore, from Eq.(\ref{diag}) we obtain the relationship:
\begin{eqnarray}
\omega_{B}({\bf{q}_\parallel},\alpha)=& \{ h^{2}-2hR^{x}J\gamma [\cos
(q_{x}a)+\qquad \qquad \\ \nonumber
&\qquad \cos (q_{y}a)+ \cos (\alpha a)]\}^\frac{1}{2}
\end{eqnarray}
where $\bf{q}_\parallel$ is a two-dimensional wave vector.

We are interested in understanding how the surface and localized
impurity modes affect each other. In order to do that, we start
our analysis by considering a film with 5 layers, where the first
layer is taken as the impure one. In this case, the $\Delta$
matrix can be expressed as:
\begin{equation}
\mathbf{\Delta }^{(5)}=\left(
\begin{array}{ccccc}
\Delta _{I} & \Delta _{IB} & 0 & 0 & 0 \\
\Delta _{IB} & 0 & 0 & 0 & 0 \\
0 & 0 & 0 & 0 & 0 \\
0 & 0 & 0 & 0 & \Delta_{S} \\
0 & 0 & 0  & \Delta _{S} & \Delta _{1}
\end{array}
\right).
\end{equation}
\begin{center}
\begin{figure}
\includegraphics[width=1.0\linewidth]{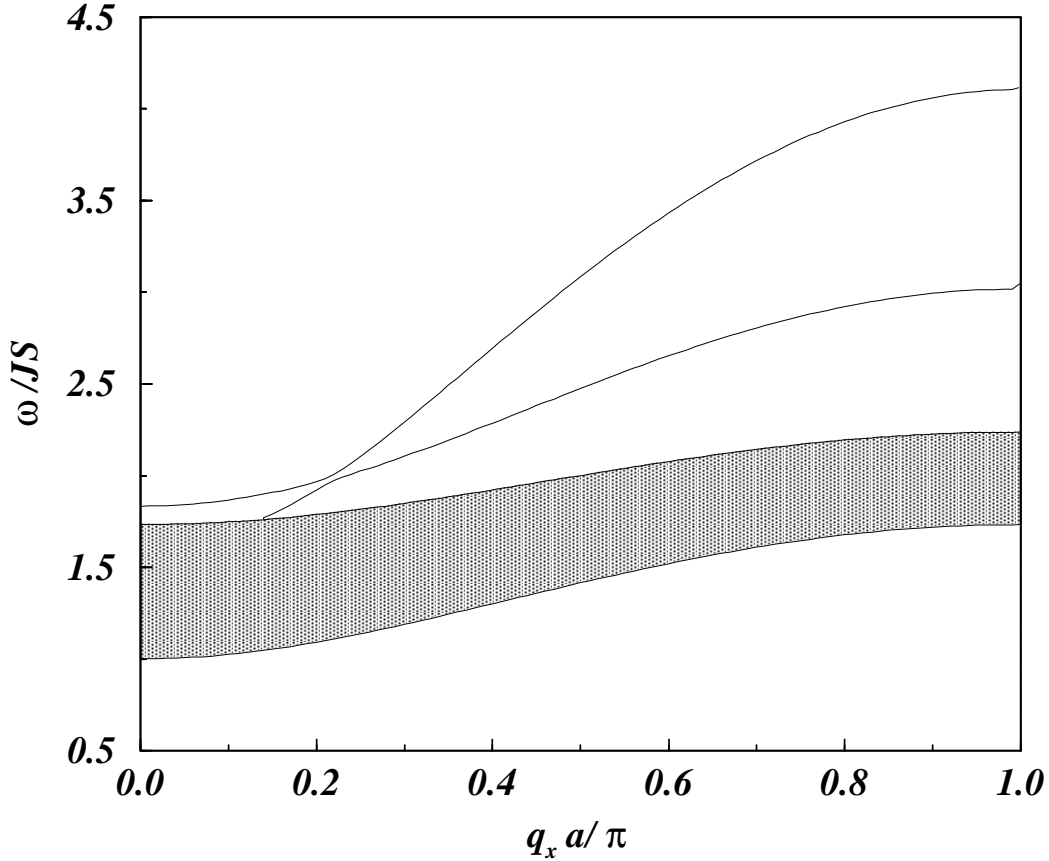}
\caption{SW dispersion for a thin ferromagnetic film with an impurity layer at the surface,
for $J_I =1.5J$. The bulk band is represented by the shaded region.
Parameters in the text.}
\end{figure}
\end{center}
Figure 1 shows a plot of frequency as a function of wave vector
$q_{x}a/\pi$, where $q_{y}=0$, for an impurity layer located on
the surface of a 5 layers film, with the parameters: $R^{x}=0.25$,
$J_{I}=1.5J$, $J_{I}^{^{\prime }}=1.25J$, $J_S=1.5J$, $h_S=1.5h$,
$h_{I}=1.35h$ and $J_{SI}= 1.0J$. The shaded area in the graph
corresponds to the region containing the volume modes of the
infinite case. One can observe the presence of two surface
branches located above the bulk region. The branches are
associated with the two surfaces of the film, one of these being
the impurity layer. The figure also shows a ``crossover'' effect
(with mode repulsion) between the surface and the surface impurity
modes at $q_x\approx 0.22$. The existence of this crossover is
dependent on the strength of the impurity exchange, as can be seen
in the next graph. The results in Fig. 2 were calculated for a
smaller value for the exchange constant in the impurity layer,
namely $J_I = 0.5J$, with the remaining parameters being the same
as those in Fig. 1. The graph shows two localized SW branches. The
lower frequency mode does not display any shift in relation to the
result in Fig. 1. On the other hand, the upper SW branch shows a
significant shift in comparison with the corresponding branch in
Fig. 1. Thus, we may associate the lower frequency branch with the
pure surface of the film, whereas the upper branch can be
associated with the surface impurity layer mode, since the shift
occurs as a direct consequence of the smaller value of $J_I$.
Also, due to the frequency shift, no mode repulsion effects are
found.

Let us now consider the situation in which the impurity layer is
located in the interior of the film. Results for this case are
shown in Fig. $3$, for a $5$ layer film, in which the impurity
layer corresponds to the $3$rd layer of the film. The $\Delta$
matrix for this configuration is:
\begin{equation}
\mathbf{\Delta }^{(5)}=\left(
\begin{array}{ccccc}
\Delta _{1} & \Delta _{S} & 0 & 0 & 0 \\
\Delta _{S} & 0 & \Delta_{IB} & 0 & 0 \\
0 & \Delta_{IB} & \Delta_{I} & \Delta_{IB} & 0 \\
0 & 0 & \Delta _{IB} & 0 & \Delta _{S} \\
0 & 0 & 0  & \Delta _{S} & \Delta _{1}
\end{array}
\right).
\end{equation}
The SW frequencies are obtained following a procedure similar to
the previous case. The parameters are $h_{I}=1.45h$, $J_{I}=1.5J$,
$J_{I}^{^{\prime }}=1.25J$, $h_{S}=1.35h$, $J_{S}=0.5J$,
$J_{S}^{^{\prime }}=1.0J$, $J_{SI}=1.0J$, with $R^x=0.25$. The
plot shows three coupled modes. For small wave vectors, there are
two degenerate frequency branches related to the surfaces, as well
as a lower frequency one (impurity mode). As the wave vector
increases, the degeneracy is lifted. In the range $0.5 \pi <q_xa
<0.7 \pi$, there are three distinct modes. That happens as a
consequence of a double crossover, due to the mixing of the
surface and impurity modes. For larger values of $q_x$, the
surface modes are again degenerate.

Figure $4$ shows a graph of the frequencies of localized SW modes
as a function of the effective field $h_I$ in the impurity layer.
We used a film with the same parameters as the one discussed in
Fig. 3. Results are shown for three different values of in-plane
wave vector, namely $q_xa=0.01 \pi$ (solid line), $q_xa=0.5 \pi$
(long-dashed line) and $q_xa=1.0 \pi$ (dotted line). In contrast
with the semi-infinite case presented in Ref. \cite{Costa}, the
graph displays a rich spectrum of localized modes. The
discontinuity in the graph is a consequence of the merging of some
of the modes with the bulk band for certain values of $h_I$. For
each value of wave vector, the graph shows two different behaviors
of the localized SW branches: some branches display a strong
dependence on $h_I$, whereas the remaining branches are not
noticeably influenced by a varying field, except at larger values
of $h_I$, where mode mixing effects are observed. These
contrasting behaviors reflect the distinct nature of these modes,
with the impurity modes being associated with the branch that is
strongly influenced by the $h_I$ field. These impurity modes have
acoustic nature for small values of $h_I$. As the effective field
is increased, the impurity mode merges with the bulk band and
eventually reappears as an optical mode. For the parameters used
in this calculation, the surface modes are found above the bulk
band.
\begin{center}
\begin{figure}
\includegraphics[width=1.0\linewidth]{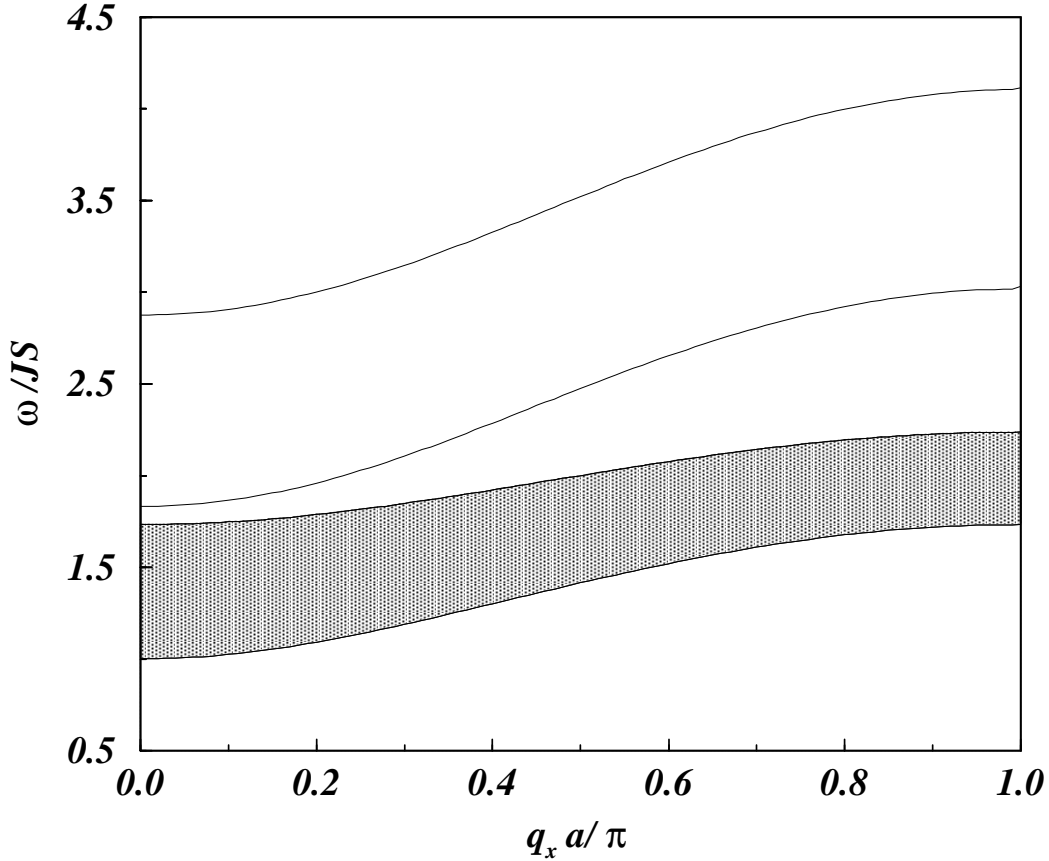}
\caption{SW dispersion, for a system similar to the film of Fig. 1,
with $J_I = 0.5J$. }
\end{figure}
\end{center}

\begin{center}
\begin{figure}
\includegraphics[width=1.0\linewidth]{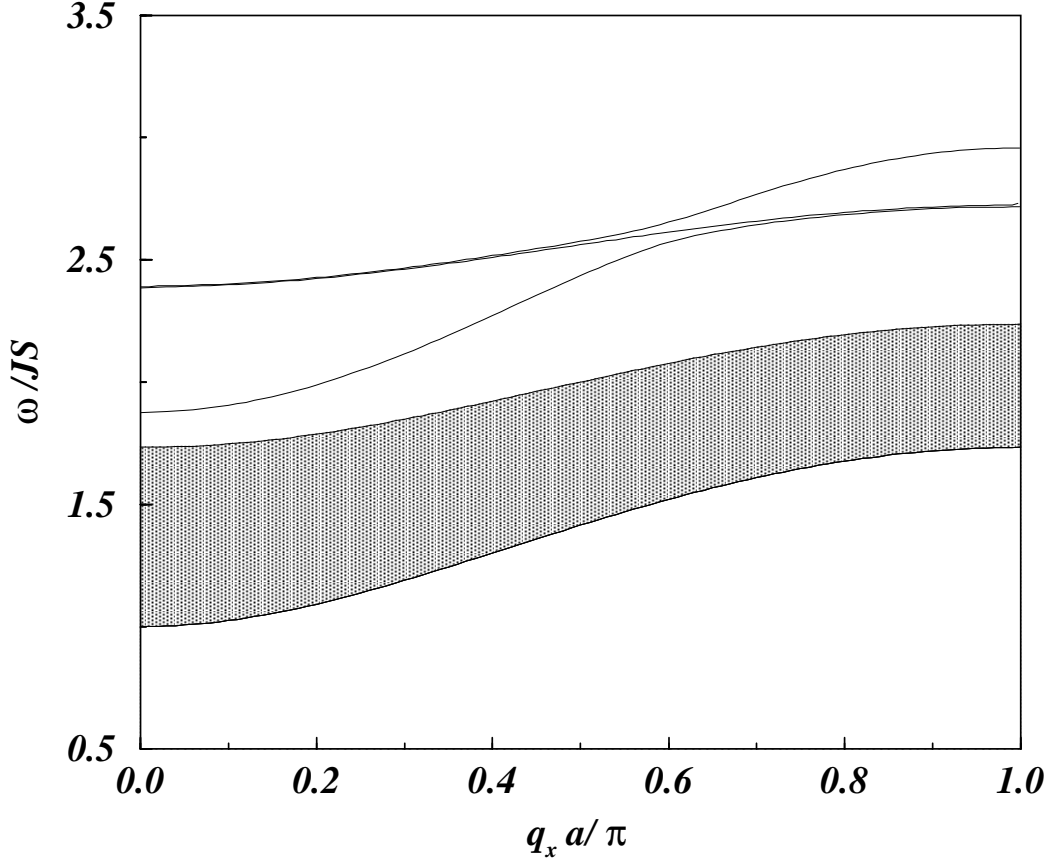}
\caption{SW dispersion, for an impurity layer in the interior of the
film,
corresponding to the third atomic layer. The bulk band
is represented by the shaded region. Parameters in the text.}
\end{figure}
\end{center}

\begin{center}
\begin{figure}
\includegraphics[width=1.0\linewidth]{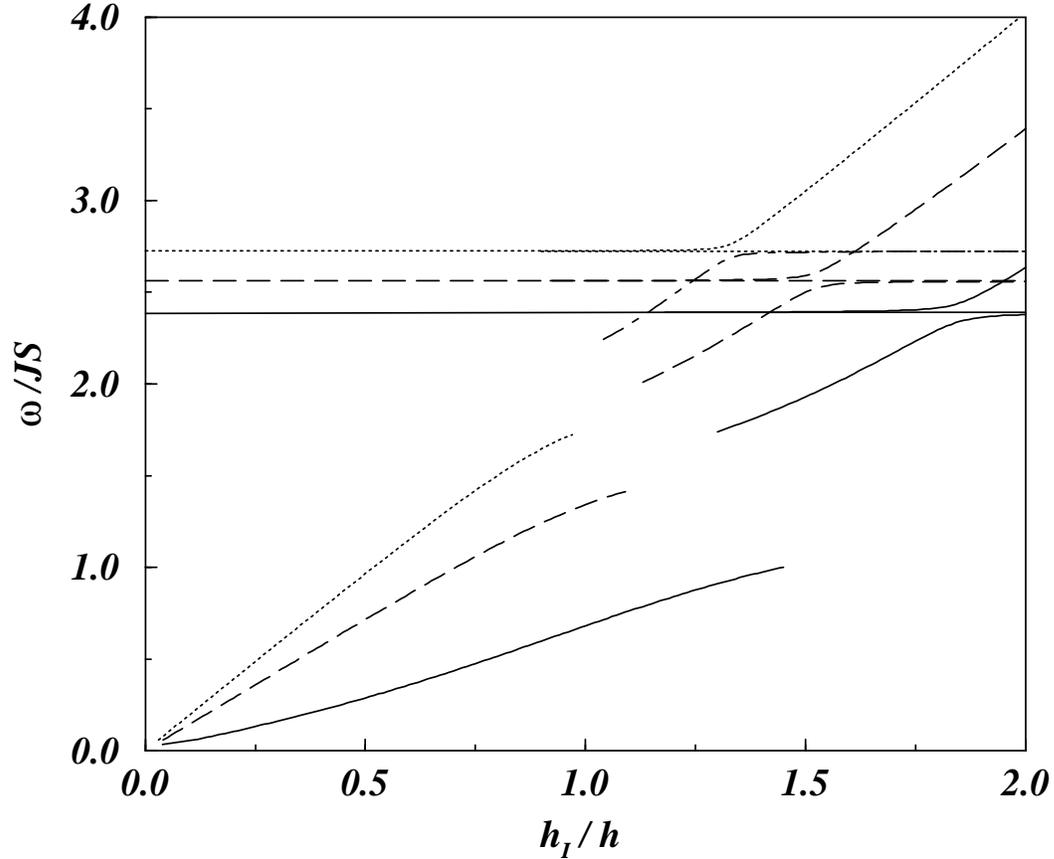}
\caption{Localized SW frequencies as a function of impurity magnetic field
for three values of wave vector. Parameters in the text.}
\end{figure}
\end{center}
\begin{center}
\begin{figure}
\includegraphics[width=1.0\linewidth]{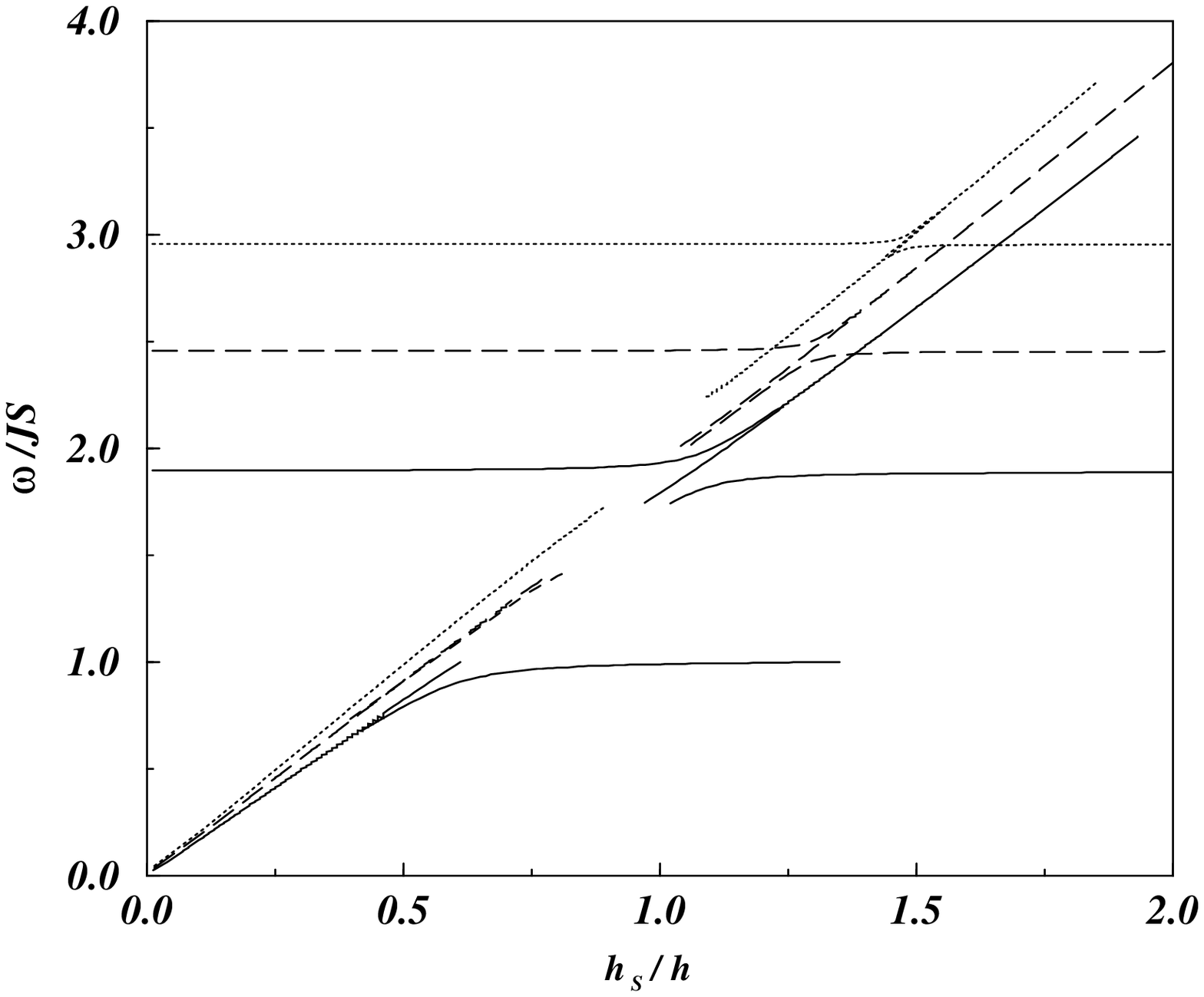}
\caption{Localized SW frequencies as a function of surface field
for three values of wave vector (solid curve, $q_xa/\pi=0.5$; long-dashed curve, $q_xa/\pi=0.01$
and  dot-dashed curve, $q_xa/\pi=1.0$). The impurity layer corresponds to
the 3rd layer of a 5 layers film.}
\end{figure}
\end{center}

This difference in behavior is also evident in the results of Fig. $5$,
which shows localized SW frequencies as a function of the effective surface field $h_S$,
for the same wave vector values as in the previous results and $h_I/h = 1.45$.
The graph shows the influence of $h_S$ on the surface modes. For small values of this
effective field, the surface modes are degenerate and have acoustic behavior. On the
other hand, the impurity modes are not affected by the variation of $h_S$. As the
field is increased, the surface modes merge with the bulk band, and an acoustic impurity
mode is observed. Mode mixing effects are also found. For larger values of $h_S$,
surface modes are found above the bulk band and there are two impurity modes: one optical
branch, and one acoustical.

\section{Conclusions}

In summary, we developed a Green's Function formalism for
calculating the SW spectrum of a ferromagnetic thin film with an
impurity layer, in the framework of the transverse Ising Model,
using the RPA to obtain equations of motion for the bulk, surface
and impurity SW modes. We calculated the Green's functions
analytically through an inhomogeneous matrix equation in a closed
form. The formalism can be applied for any temperature, and we
obtained explicit solutions for $T\geq T_{c}$. The results showed
the effect of the exchange coupling between spin sites in the
impurity layer on the localized modes of a film in which the
impurity layers correspond to one of the surfaces. We also
obtained results for the effect of the position of the impurity
layer in the film. We found that the placement of the impure layer
has a strong effect on the spectrum of localized modes. In all
cases, we observed strong mode mixing effects between localized
impurity and surface modes. In addition, we calculated the
localized SW frequencies as functions of the effective fields
$h_I$ and $h_S$. This allowed us to analyze the nature of the
modes, by observing the response of the frequency branches to the
variation of the effective fields. In the limit $N\rightarrow
\infty$, the results agree with previous calculations for
semi-infinite systems \cite{Costa}. This formalism can also be
applied to obtain the spectral intensities of the bulk and
localized SW modes. Currently, we are extending the formalism in
order to include the effects of localized impurities in thin
films.

The authors gratefully acknowledge partial support the Brazilian
agency \textbf{CNPq}.


\end{document}